# A quick test of the WEP enabled by a sounding rocket

Robert D. Reasenberg,[1] Enrico C. Lorenzini,[2] Biju R. Patla,[1] James D. Phillips,[1] Eugeniu E. Popescu,[1] Emanuele Rocco,[1,3] and Rajesh Thapa[1]

[1]Smithsonian Astrophysical Observatory, Harvard-Smithsonian Center for Astrophysics, 60 Garden St., Cambridge, MA 02138.

[2]Faculty of Engineering, Universita di Padova, Padova I-35122, Italy.

[3]Current address:

E-mail: reasenberg@cfa.harvard.edu

Abstract. We describe SR-POEM, a Galilean test of the weak equivalence principle, which is to be conducted during the free fall portion of a sounding rocket flight. This test of a single pair of substances is aimed at a measurement uncertainty of $\sigma(\eta) < 10^{-16}$ after averaging the results of eight separate drops, each of 40 s duration. The weak equivalence principle measurement is made with a set of four laser gauges that are expected to achieve 0.1 pm $Hz^{-1/2}$. We address the two sources of systematic error that are currently of greatest concern, magnetic force and electrostatic (patch effect) force on the test mass assemblies. The discovery of a violation ($\eta \neq 0$) would have profound implications for physics, astrophysics and cosmology.

## 1. Introduction

The weak equivalence principle (WEP) underlies general relativity, but violations are possible in most theories being developed to unify gravity with the other forces. A WEP violation is characterized by $\eta$, which is identically zero in any metric theory of gravity, including general relativity.

$$\eta_{AB} = \frac{a_A - a_B}{(a_A + a_B)/2} \qquad (1)$$

[e.g., Fischbach and Talmadge 1999] where $a_A$ and $a_B$ are the acceleration of bodies A and B, these bodies are moving under the influence of identical gravity fields, and there is no other cause of the acceleration. The discovery of a WEP violation would have profound implications for physics, astrophysics and cosmology.

The present best tests of the WEP are made using a rotating torsion pendulum and yield $\sigma(\eta) = 1.8\ 10^{-13}$ [Schlamminger et al. 2008]. Advancement of this approach has slowed because of intrinsic problems with the suspension fiber that may be overcome by operating at LHe temperature [Newman 2001, Berg et al. 2005]. Torsion balances at LHe temperature are also being used to study short range gravity [Hammond 2007]. There are several proposals for better WEP tests in an Earth-orbiting spacecraft [Sumner 2007, Nobili et al. 2009, Chhun et al. 2007]. The most sensitive among these is the satellite test of the equivalence principle (STEP) [Overduin et al. 2009]. This cryogenic experiment is based on technological heritage from the successful GPB mission [GP-B]. STEP is aimed at a measurement uncertainty of $\sigma(\eta) < 10^{-18}$.

## 2. Instrument Design and Operation

The sounding-rocket based principle of equivalence measurement (SR-POEM) will achieve a sensitivity $\sigma(\eta)<10^{-16}$ in the ≈ 10 minute science portion of a single flight of a suborbital rocket [Reasenberg and Phillips, 2010]. Here, we start by briefly describing SR-POEM in terms of the present version of its evolving design, with emphasis on recent changes. We then address the six critical areas where technology development is required and the approaches we are taking in the most problematic.

In the sounding rocket payload, we compare the rate of fall of two test mass assemblies, TMA-A and TMA-B, which contain test substances, "A" and "B," respectively. Each TMA comprises a pair of cubes connected by one or two bars or tubes. Each cube is surmounted by a flat mirror with a gold surface. The selection of test substances has been discussed in many papers [e.g., Blaser 1996]. The nominal substance pair for the first flight is aluminum and lead. In one TMA, cylinders of aluminum are removed from the cubes and replaced by tubes of lead without changing the TMA mass and having minimal effect on the moments of inertia. About half of the mass will be lead. The cubes are arranged in a square lying in a plane perpendicular to the fore-aft axis of the payload (the z axis) and close to the CM of the free-flying payload. The two (joined) cubes along a diagonal of the square contain the same test substance.

At a distance of 0.3 m along the z axis from the CM, there is a highly stable ULE glass plate that holds four (concave) cavity entrance mirrors for our tracking frequency laser gauges (TFG) [Reasenberg et al. 1995, Phillips and Reasenberg 2005]. These mirrors have the same in-plane spacing and orientation as the cubes. By design, these mirrors have about the same reflectivity as gold $(97.5\pm1\%)$ at our laser wavelength of 1550 nm. Each gold mirror on a cube forms a measurement cavity with the corresponding concave mirror on the TFG plate. In operation, each cube is continuously observed by the corresponding TFG.

The cubes are surrounded by the plates of a multi-component capacitance gauge that measures TMA motion in all degrees of freedom. The capacitance gauge plates can be used to apply force or torque electrostatically. This capability, when combined with the sensing, is called the TMA suspension system (TMA-SS), which is used both for high force applications – initial capture and slewing – and for fine placement.

When the sounding rocket payload reaches an altitude of about 800 km, the science-measurement phase of the flight begins. By that time, the TMA have been uncaged, the z axis has been aligned with the mean nadir direction of the pending "drop," the TMA charges have been neutralized, and the TMA-SS has been used to assess and then correct the position and motion of each TMA in all six degrees of freedom. Just before the start of a drop, the TMA-SS is shifted to its "passive mode" in which there is no driving force applied and excitation voltage for capacitance gauge position readings is reduced. The attitude control system will also likely be in passive mode during the measurements and part of the setup.

With the preparation completed, data taken by the TFG may be used for a WEP measurement lasting $Q$ (= 40 s, see below). There will be eight such measurements during the science-measurement phase of the flight, with the instrument orientation reversed between successive measurements to cancel most systematic errors. This "payload inversion" has the effect of reversing the WEP violating signal in the instrument coordinate system. During the payload-inverting rotational slews, the TMA-SS again controls each TMA's six degrees of freedom. The nominal apogee of 1200 km provides adequate time for the eight drops and seven inversions.

Finally, TFG data from the measurement cycles will be combined in a weighted-least-squares fit to estimate $\eta$ and its formal uncertainty. Required auxiliary data will include the payload trajectory, payload attitude from the spacecraft attitude control system, TMA transverse velocities from the capacitance gauge measurements made just before and just after the WEP measurement of duration Q, and an Earth gravity model.

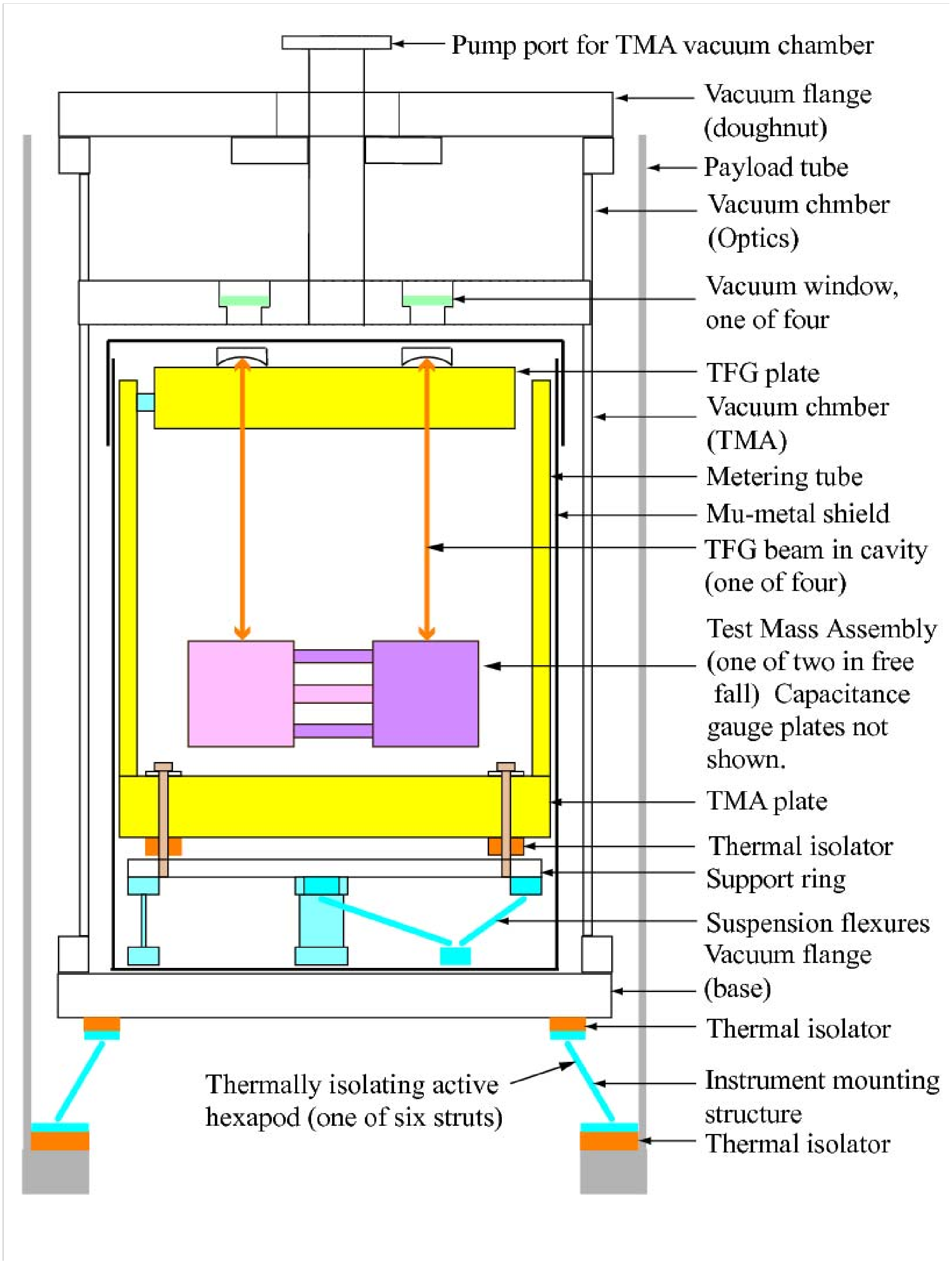

**Figure 1.** Precision instrument inside vacuum chamber inside 14 inch payload tube. Not shown here are the vacuum port for the optics chamber, the capacitance gauge electrode sets, and the TFG optics.

*2.1. EP measurement sensitivity.*

If a single laser gauge has distance measurement uncertainty $\sigma_0$ for a averaging time $\tau_0$, then the uncertainty in the estimate of acceleration, based on many measurements (assuming white noise) with total time *T*, is

$$\sigma_{acc} = \frac{\sigma_0(\tau_0)}{Q^2}\sqrt{\frac{\tau_0}{T}}K \tag{2}$$

where *Q* is the free-fall time, and $K = 12\sqrt{5} \approx 27$, assuming that position and velocity are also estimated. The corresponding WEP sensitivity for the pair of TMA and four TFG is $\sigma(\eta) = \sigma_{acc}/(R\,g(h))$, where *R* is the fraction of TMA mass that is test substance, *g(h)* is the gravitational acceleration, and *h* is the altitude of the instrument. In a reasonable scenario, $\sigma_o = 0.1\text{pm}$ (for $\tau_0 = 1s$), *Q* = 40 s, *R* = 0.5, and there are eight measurements so that *T* = 320 s. This yields $\sigma(\eta) = 0.2 \times 10^{-16}$ before allowance for uncorrected sources of systematic error.

*2.2. Instrument Design.*

The principal driver of the instrument design is the suppression of systematic error. The CM positions of the two TMA are nominally the same, as described below. The z coordinate of a TMA CM is obtained from the average of the two TFG measurements made between the mirrors on the two cubes and the corresponding concave mirrors mounted on the TFG plate.

The instrument precision structure is in the TMA chamber, Fig. 1. It is constructed of ULE glass, which has a low coefficient of thermal expansion (CTE, about $10^{-8}$ / K). This structure has three major components: 1) The TMA plate holds capacitance gauge electrode sets and provides access for the TMA caging mechanism. 2) The metering structure provides a rigid, stable connection between the TMA plate and the TFG plate. 3) The TFG plate supports the four TFG end mirrors, which are mounted over holes in the plate to allow a view of the TMA. The associated beam steering devices and the optical detectors are housed in the optics chamber, where we can use selected plastic insulation and vacuum grease lubricant, since the pressure need be no lower than $10^{-4}$ Torr.

The TFG measurement cavities are inside the TMA chamber, which is pumped down, baked and allowed to cool before launch. A small pump runs during the flight to maintain a pressure of $10^{-10}$ Torr. The prime candidates are an ion pump and a sorption pump. All electrical and optical connections to the instrument are made through the end flanges of the TMA and optics chambers (bottom and top in the Figure).

The physics package, i.e., the dual vacuum chamber and its contents, is connected to the payload tube through a thermally isolating hexapod motion system (Stewart platform). By means of the control of the hexapod, the experiment becomes effectively drag free and significant reduction is achieved in its rotation and translation with respect to the inertial frame defined by the TMA. On a low-drag trajectory, the Stewart platform actuators do not need a large range.

*2.3. Laser gauge.*

The principal measurements are made by the four TFG observing the four cubes that are joined pairwise to form the two TMA. In operation, each cube is continuously observed. Because of its high precision, 0.1 pm in 1 s, the TFG allows a high accuracy test of the WEP during the brief period afforded by a sounding rocket. The TFG speed of measurement, coupled with the payload inversions, shifts upward the frequency band of interest for the measurements to 0.007 Hz and above. This shift reduces susceptibility to systematic error.

The SR-POEM measurement will employ four Fabry-Perot resonators, each with a curved mirror at the TFG plate and a plane mirror at one of the TMA cubes. The optical beam will be injected through the curved mirror and kept aligned by an automated system that was derived from the cavity alignment system demonstrated by Sampas and Anderson [1990].

Our current laser gauge implementation (Fig. 2) is the Semiconductor Laser TFG (SL-TFG). In it, the Variable Frequency Source (VFS) comprises a tunable distributed feedback (DFB) semiconductor laser operating at 1550 nm. The VFS' optical frequency is locked to the cavity formed by the mirrors. For all four TFG's, there is one reference laser, which is locked to a very stable cavity or to an atomic line. A portion of the light from each TFG's tunable laser is heterodyned against a portion of the reference laser light to provide the radio frequency output whose changes equal changes in optical frequency difference. The WEP observable is the difference of the averaged outputs from TMA-A and TMA-B. The SR-POEM configuration has four SL-TFG's with closely-matched paths. Because the measured paths are nearly equal, the stability of the reference laser frequency is relatively noncritical.

### *2.4. Beam alignment and TMA rotation.*

Even if the TMA are not rotating with respect to the mean orientation of the payload, the $\approx 10^{-6}$ radian orientation variations of the payload will cause a misalignment of the optical cavity in the absence of an alignment servo. We address this at two levels. First, the alignment servo reacts quickly but results in "beam walk" on the curved mirror of $\approx 0.1$ μm, which should be compared to the spot size, $\omega = 0.7$ mm. Second, TFG and capacitance gauge readings are used to drive the Payload Servo, a hexapod motion system that holds constant the relative position and orientation of the experiment and TMA.[1] The alignment system is required for countering higher-frequency disturbances of low amplitude, while the Payload Servo minimizes variations of the positions of the TMA with respect to their immediate surroundings.

Before each drop, the payload is oriented and the TMA-SS used to set the initial conditions, which ideally would be: the CM of the two TMA collocated and comoving with the payload CM; the physics package inertially pointed; and the TMA oriented toward the TFG and not rotating with respect to the rest of the physics package. The beam alignment system provides ancillary information about the relative orientation of the experiment and TMA by monitoring the position on the input mirror where the TFG beam is injected.

### *2.6. Thermal design and analysis*

In the short ($\approx 10^3$ s) period of free fall, it will not be possible to reach thermal equilibrium. Yet, thermal stability is an essential aspect of the WEP experiment. Our approach is to ensure long thermal time constants and, where necessary, to use low CTE materials. By flying near midnight, we avoid direct solar heating,[2] which would greatly exceed other thermal perturbations and be highly directional in spacecraft coordinates.

We seek to minimize two adverse thermal effects. First, temperature change within the precision measurement system can result in direct measurement error. For example, it could cause a distortion of the TFG plate. Second, a temperature change in any part of the payload can result in a change of the mass distribution (e.g., through the expansion of the structure) and therefore a change in the local gravity, with different changes in the accelerations of the two TMA. Note that, in both cases, the perturbation would need to be synchronous with the measurement-inversion cycle to affect the WEP test.

---

[1] The drive signal for the servo is based on a weighted average of the data for the two TMA. The weights are specific to each degree of freedom.

[2] Even at midnight, the Sun can be seen at the summer solstice (a few degrees above the horizon to the north) from 1200 km above the Wallops Flight Facility.

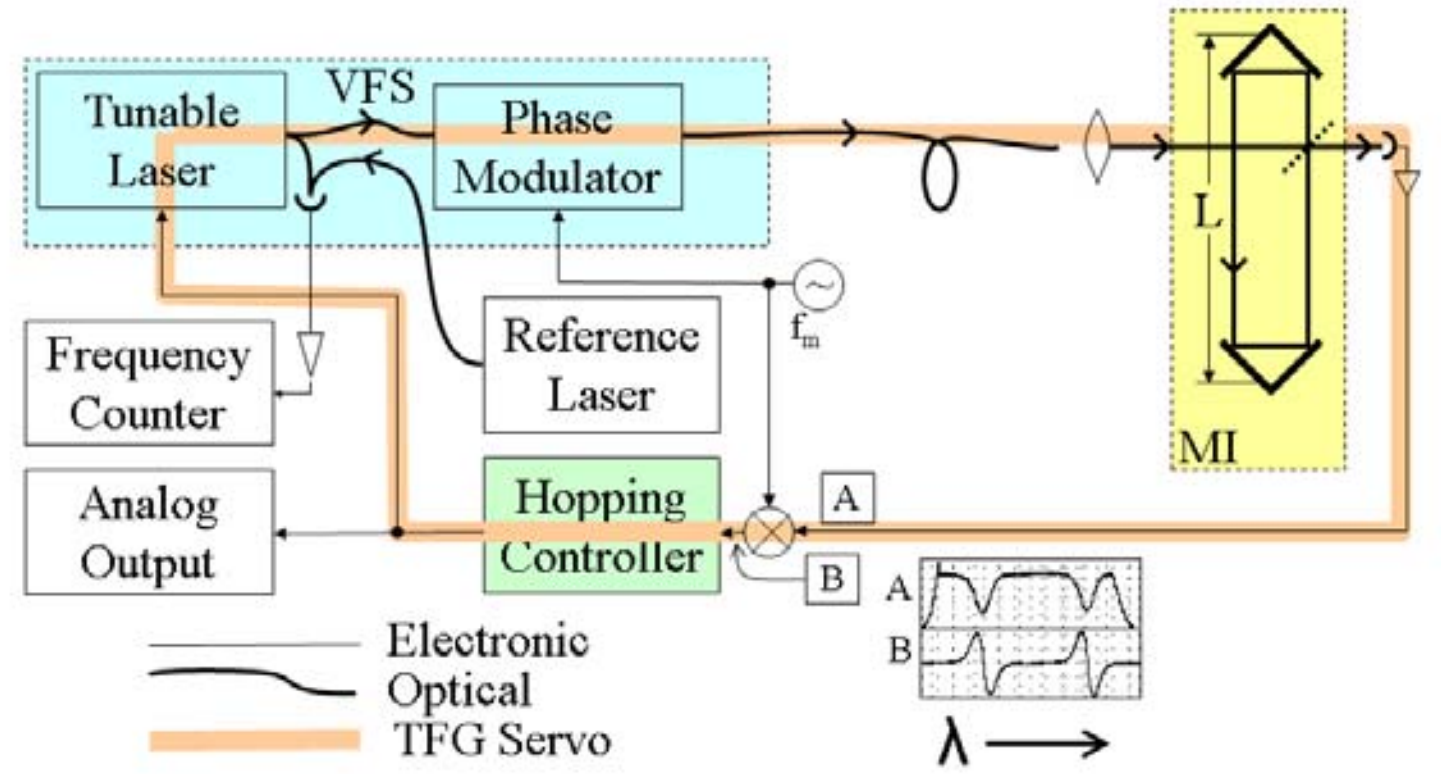

**Figure 2.** Block diagram of the SL-TFG. Its servo loop is closed by Pound-Drever-Hall locking, with a unity-gain frequency ~1 MHz. The insert shows signals at marked points in the diagram as the VFS is swept over two fringes of a low finesse cavity. A, resonant dip. B, dispersion curve derived from modulation and synchronous detection.

Our approach to minimizing the thermal effects is illustrated in Fig. 1. An aluminum vacuum chamber with ¼ inch wall, containing the precision measurement system, is mounted to the payload tube through a thermally isolating active hexapod structure. The outside of the chamber and payload tube are gold plated giving those surfaces a low emissivity, $\varepsilon = 0.02$. Radiative exchange between the payload tube ($\varepsilon = 0.1$) and the chamber wall causes the chamber to approach the tube temperature with a time constant of $1.5 \times 10^5$ s. Inside the vacuum chamber, the most critical component is the TFG plate, which is a ULE disc of 20 cm diameter and 4 cm thickness with $\varepsilon \approx 1$. It is suspended from the metering tube by three titanium flexures. Radiative exchange between the chamber wall ($\varepsilon = 0.1$) and the TFG plate causes the plate to approach the chamber wall temperature with a time constant greater than $10^5$ s. The corresponding time constant for the much thinner (1 cm) and more exposed metering tube is $2.8 \times 10^4$ s. These time constants were calculated without consideration of the beneficial (thermal shielding) effect of the Mu-metal magnetic shield between the chamber and the ULE structure.

## 3. Technology Issues

As previously noted, there are six technology issues that we view as critical to the success of the SR-POEM Mission. In approximate descending order of concern they are: Magnetic force, Electrostatic force, Uncaging, Reliability, Local gravity field, and Laser gauge measurement precision. Here we discuss the first four of these. The problems associated with the local gravity field are solved by the inversions and the movement of the Stewart platform. The laser gauge development is proceeding well.

### *3.1 Magnetic force*

The magnetic force on a TMA depends on its magnetic moment and the gradient of the magnetic field.

$$F_z = \sum m_i \frac{\partial B_i}{\partial z}, \quad \text{where} \quad i = \{x, y, z\} \quad \text{and} \quad \vec{m} = \vec{m}_p + \vec{m}_{in} \qquad (3)$$

In Eq 3, $m_p$ and $m_{in}$ are the permanent and induced magnetic moments. Further, on payload inversion, only some of the components of magnetic force invert and thus cancel. The current design has a single magnetic shield of thickness, $t_{sh} = 3$ mm, and radius, $r_{sh} = 0.12$ m. We assume the shield to be made of Mu-metal and take its permeability to be $\mu = 5\times10^4$. We estimate crudely that it reduces the external field by $\mu t_{sh} / r_{sh}$, which yields a factor of about $10^3$. Similarly, it produces a gradient of order $B_{lk} / r_{sh}$, where $B_{lk}$ is the field leaking through the shield. (We are currently making a more realistic estimate of both of these quantities by applying finite element modeling to our shielding geometry.)

We estimate that the TMA will have a permanent magnetic moment, $m_p \approx 2\times10^{-8}\,\mathrm{A\,m^2}$ based on Su et al. [1994]. (Basing the estimate on Mester and Lockhart [1996] yields a higher moment, but this work was on very small samples measured at 4-30°K.) Thus, the magnetic force will produce an acceleration of $3\times10^{-15}$ g(h), and this needs to be reduced more than 30-fold to meet the Mission goal. To address this problem, we are investigating a second shield. The leading candidate is an open ended shield, which will be most effective on the transverse component of the external field. This could be augmented by a canceling coil to null the axial field. We expect the second shield to further reduce $B_{lk}$ by at least 100 fold.

### *3.2 Electrostatic force*

As the TMA is freed from its launch restraint, it will acquire a potential of a few × 0.1 V. This is easily measured by briefly applying a constant potential to the capacitance gauge electrodes and using the TFG to determine the acceleration. We plan to neutralize the TMA iteratively by illuminating the TMA and one of the surrounding (TMA-SS) electrodes with a UV LED, and biasing the charge flow by putting a potential on the illuminated electrode [Sun 2009]. The remaining electric field, due to the average surface potential of the TMA top and bottom faces, will be made nearly zero by applying the same potential to the corresponding TMA-SS electrodes.

The more difficult problem comes from the spacial variation of potential across a surface, often called the patch effect [see Camp et al. 1991]. A standard solution is to use a gold surface, and to ensure that it is clean. The gold is often deposited over another thin film of material that will make the gold deposit more uniformly, e.g., germanium. An extensive Kelvin-probe study by Robertson et al. [2006] examined numerous surface-coating combinations, finding spacial fluctuations of potential of 1 to 2 mV rms with respect to the mean on the "best" surfaces. These results were at the discreetization level of their Kelvin probe. We are investigating the use of a more sensitive Kelvin probe with higher special resolution and much smaller discreetization.

Recent work by Pollack et al. [2008] looks directly at the force between a pair of surfaces using the high sensitivity of a torsion balance. They find that for frequencies above 0.1 mHz there is a white potential-fluctuation spectrum with density $30\,\mu\mathrm{V}/\sqrt{Hz}$. For the case of SR-POEM, the four 140 s measurement cycles (each containing two 40 s measurements and two inversions) yield a fluctuation of 1.3 μV For comparison, a fluctuation of 200 μV yields an error of $10^{-17}$ g(h), which would be completely acceptable.

### *3.3 Uncaging*

The TMA need to be held firmly ("caged") during launch because of the high levels of vibration and upward acceleration from the engines. Since the experiment requires that the TMA be in a UHV environment, the launch supports ("caging pins") and the TMA are expected to stick together as we attempt to uncage. LISA has approached this problem with a three-stage uncaging system [Jennrich 2009]. We are investigating the use of a non-stick surface on the ends of the caging pins made by the self assembly of a monolayer of mercaptan (i.e., R-S-H). Even after being subjected to a pressure sufficiently large to cause plastic deformation of the gold, such a surface has been shown not to stick by Thomas et al. [1993], who reports on the use of *n*-docosanethiol, $CH_3(CH_2)_{21}SH$. Contact between the caging pins and the TMA would be at the bottom of small holes in the TMA so as to minimize the electrostatic force due to surface potential changes caused by the pins.

*3.4 Reliability*

A sounding rocket experiment is inevitably carried out in a short period of time. It must work as launched without “tuning up” by commands from a ground station. Algorithms must be robust, not necessarily optimal. The flight software must take in stride all anomalies and yet take data on schedule. This is a challenge given the complexity of the SR-POEM event sequence, which includes pre-measurement calibration and set up, alternating measurement and payload inversion and post-measurement recalibration. The control program (sequencer) for the mission will need to undergo extensive and realistic simulation. In addition, we plan a series of tests of the physics package onboard a “zero-g” airplane flight. This will be used to test key aspects of the mission such as uncaging, charge neutralization, the operation of the payload servo, and flight software.

## 4. Conclusion

We are developing a Galilean test of the WEP, to be conducted during the free-fall portion of a sounding rocket flight. The test of a single pair of substances is aimed at a measurement uncertainty of $\sigma(\eta)<10^{-16}$ after averaging the results of eight separate drops during one flight. We have investigated sources of systematic error and find that all can be held to well below the intended experiment accuracy. A detailed error budget is in preparation and will be published soon [Patla et al. in preparation].

At the core of the mission design is the mitigation of systematic error, which is supported by the three stages of measurement differencing. First, the TMA positions are each measured by laser gauges with respect to the commoving instrument. These measurements contain the acceleration of the instrument and initial velocity errors, reduced by the payload servo (hexapod). The change in the length of the measurement path over the 40 s drop period is at the sub-micron level. Second, those measurements are differenced, removing the residual instrument acceleration. Third, the payload is inverted and the differential accelerations are differenced. This last step adds the WEP violating signals and subtracts the accelerations of the TMA due to fixed electrostatic, gas pressure and radiation pressure forces, the fixed comoponents of the magnetic force, and to local gravity and the symmetric part of the Earth’s gravity gradient. (The next, asymmetric term is both easily calculated and below the threshold of interest.) Ideally, at the start of a drop, the CM of each TMA should be at the CM of the payload. There will be an error in this setting due to the uncertainty in the on-board calibration, which will be much better than the corresponding ground-based measurement. However, the initial conditions for the drop will be highly reproducible because the key position (along the z axis) is sensed by the laser gauge.

Before we can perform a successful WEP experiment, we will need to develop an enhanced version of our TFG laser gauge, which will have a precision of 0.1 pm $Hz^{-1/2}$ and be self aligning. If the total Mission error is to be $\sigma(\eta)<10^{-16}$, we will need to show that the magnetic and electrostatic (including patch effect) forces on the TMA do not result in systematic error larger than about $0.3 \times 10^{-16}$, each. We will also need to demonstrate that we can release the TMA after launch and control them both during the inversion slews and in preparation for the drops. Methods for achieving these goals have been identified and work has been started on the demonstrations. The use of an active hexapod connection between the payload tube and the experiment chamber solves some problems related to the mitigation of systematic error, particularly related to gravity forces from local matter. Although preliminary analysis showed no serious problems with the use of an active hexapod, this approach has not yet been subjected to detailed analysis.

A successful SR-POEM Mission would provide a 1000 fold increase in the accuracy of our knowledge of the WEP violation parameter, $\eta$. Either the discovery of a violation of the WEP or its deeper confirmation would inform the development of gravitational physics. Extension of the current work to higher accuracy appears possible by using a sounding rocket with a higher apogee. We are investigating a version of SR-POEM aimed at $\sigma(\eta)<10^{-17}$.

## Acknowledgements

This work was supported in part by the NASA SMD through grants NNC04GB30G (past) and NNX07AI11G and NNX08AO04G (present). Additional support came from the Smithsonian. We thank the staff at the Wallops Flight Facility for their generous contribution to our understanding of the capabilities and constraints of a sounding rocket flight.